\documentstyle[twoside,fleqn,espcrc2,psfig]{article}

\title{Ginzburg Criterion for the Chiral Transition}
\author{M.A. Stephanov\address{Institute for Theoretical Physics,
SUNY, Stony Brook, NY 11794-3840}}

\begin{document}
\begin{abstract}
This report is based on the work done together with J.B. Kogut and
C.G. Strouthos.  
We study a Yukawa theory with spontaneous chiral symmetry breaking and
with a large number $N$ of fermions near the finite temperature phase
transition. Critical properties in such a system can be described by
the mean field theory very close to the transition point. We show that
the width of the window of non-trivial scaling is
suppressed by a certain power of 1/N. Our Monte Carlo simulations
confirm these analytical results. We discuss implications for the
chiral phase transition in QCD.
\end{abstract}
\maketitle

\section{Introduction}

Global symmetries provide important information
about the properties of quantum field theories. 
In QCD with 2 massless quarks the global symmetries form the group: 
SU$_A$(2)$\times$SU(2)$_V$ $\times$U$_B$(1).
This fact, together with the ideas of dimensional reduction and
universality gives us a prediction for the
long-scale behavior at the chiral symmetry restoration transition
in QCD \cite{PiWi84}.

The idea of dimensional reduction is based on the observation that at
the second order phase transition a certain correlation length in the
system diverges and eventually becomes longer than the extent of the
Euclidean time.  The long-wavelength fluctuations are then
squeezed to a $d-1$ dimensional ``pancake''.
The universality tells us that the long-wavelength
behavior of a system depends on the global symmetry and the
dimensionality of the system. In particular, for the QCD with 2
massless quarks one should find O(4)$=$SU(2)$\times$SU(2) exponents at
the chiral phase transition \cite{PiWi84}.

Such an argument, based on universality, does not tell us
how this universal critical behavior sets in. To answer this
question one has to study the dynamics. In this work we
study how this universal behavior sets in for theories with large
number, $N$, of fermion species.

The fermionic fields, unlike the bosonic, do not survive dimensional
reduction. One way to see this is to recall that fermion fields
have antiperiodic boundary conditions in the time direction, which
suppress fluctuations when their correlation 
length exceeds $1/T$. In another language, the effective
$d-1$-dimensional theory for the bosonic fields near $T_c$
is a classical statistical theory. Fermion fields do not have
a classical limit \cite{St95}.

However, the fermions may affect the way that the universal behavior
sets in.
Several authors observed \cite{RoSp94,KoKo94} 
that in the large $N$
limit the exponents of the finite $T$ chiral transition in the Yukawa
model are given by the mean field (MF) theory. 
Monte Carlo simulations confirmed this result \cite{KoKo94}. 
How does it reconcile
with the universality arguments? The answer is that
the large $N$ description has its region of applicability.
The scaling for the correlation length in the window between
$1/T$ and $N^x/T$, $x>0$, is indeed given by MF. However,
when the correlation length exceeds $N^x/T$ the universal scaling
sets in.

\section{The model}

To derive the crossover exponent $x$ we study \cite{KoSt98}
a Yukawa theory with large number $N$ of fermion
species. It can be thought of as an effective description of degrees
of freedom in QCD which participate in chiral symmetry breaking.
We choose the symmetry group Z(2) instead of O(4) to simplify the
argument.
The theory is defined by the Lagrangian:
\begin{equation}\label{yukawa}
{\cal L} = {1\over2}(\partial\phi)^2 + {1\over2}\mu^2\phi^2
+ \lambda\phi^4  + \sum_{f=1}^{N}\bar\psi_f \left( 
\partial\hskip -.5em / + {g\phi}
\right) \psi_f
\end{equation}
and is regularized by some momentum cutoff, $\Lambda$.
 There are two other
important scales in the theory: the temperature, $T$, and
the mass $m$
of the thermal excitations of the scalar field. $m$ vanishes at $T_c$. 
Therefore, near the finite temperature phase transition we have 
the following hierarchy of scales: $\Lambda\gg T \gg m$.

The renormalization group (RG) evolution
of the quartic
self-coupling of the scalar field, $\lambda$, from
the scale $\Lambda$ to the scale $T$ is governed by the
RG equations of the $d$-dimensional quantum Yukawa model. 
After that, at the scale of
$T$, we pass through a crossover region: the fermions 
and nonzero Matsubara modes of the scalar fields
decouple. The evolution
below $T$ is governed by the RG equations of the scalar $\phi^4$
theory in $d-1$ dimensions.

If the window of scales between $\Lambda$ and $T$ is wide 
the renormalized coupling $\lambda$ at the scale $T$,
$\lambda(T)$, is close to the infrared fixed point of the
$d$-dimensional Yukawa theory. In the large-$N$ limit:
\begin{equation}\label{lambda}
\lambda(T) \sim {(4-d)\ T^{4-d}\over N} \quad\mbox{for} \quad 2<d<4.
\end{equation}
The case $d=4$ is special. The infrared fixed point is trivial
and is approached logarithmically as $\Lambda/T\to\infty$:
$\lambda(T) \sim 1/ [N\ln(\Lambda/T)]$.
For large $N$ this coupling is small. 
As we shall see shortly, this is the reason why the non-MF critical region
is reached only very close to the phase transition.

The quantitative relation between the size of the
non-MF  critical region, the Ginzburg region, and certain
parameters of a given system is known as the  Ginzburg criterion.
In superconductors such a parameter is a small ratio $T/E_F$,
i.e., the width of the Ginzburg region is suppressed by a power
of this parameter. In a field
theory with large number of fermions, such as (\ref{yukawa}),
such a parameter is $1/N$.

The MF approximation
breaks down because of self-inconsistency 
when the fluctuations become large.
The size
of the corrections to the MF is determined by the value
of the effective self-coupling of the scalar field.
Since the dimensionality of the coupling 
of the $d-1$-dimensional scalar theory,
$\lambda_{d-1}=T\lambda(T)$, is $5-d$,
the MF approximation breaks down
when $\lambda_{d-1} \sim m^{5-d}$. From this, 
and with the help of (\ref{lambda}),
one obtains the following criterion for
the applicability of the MF scaling:

\begin{equation}
m \gg {T\over N^x} \ , \quad x={1\over5-d} \ .
\label{ginzburg}
\end{equation}
In the special case of $d=4$ one finds:
$m \gg T/ [N\ln(\Lambda/T)]$.

The Ginzburg criterion (\ref{ginzburg}) says 
that for masses $m$ inside the window
$T \gg m \gg T/N^x$ the MF scaling holds,
while for smaller masses $m \ll T/N^x$ (i.e., closer to the transition) 
the non-trivial $d-1$ Ising scaling sets in. We see that the size
of this latter, non-trivial critical region is suppressed at large
$N$.%

\section{Lattice}

The strategy we use to confirm our analytical results 
is the following. First, we show that the
behavior at sufficiently large correlation length is given by the
universality arguments, which in this case predict Ising $Z(2)$ critical
exponents. Second, we identify the boundary of the MF region,
test the scaling and extract the exponent $x$.

We discretize the $d=3$ theory (\ref{yukawa})
on a cubic $L_s^2\times L_t$ lattice in the
following standard way \cite{HaKoKo93}:
\begin{eqnarray}\label{yukawa-lat}
S &=& {\beta N\over4}  \sum_{{\tilde{x}}}
\phi^{2}_{\tilde{x}}
+\sum_{i=1}^{N/2} \Big( 
\sum_{x,y} \bar{\chi}^{i}_{x} M_{x,y} \chi^{i}_{y}
\nonumber\\ &&
+{1\over8} \sum_{x} \bar{\chi}^{i}_{x} \chi^{i}_{x} 
\sum_{ \langle \tilde{x},x \rangle} \phi_{\tilde{x}} \Big) .
\end{eqnarray}

\subsection{The FSS results}

In order to study the critical behavior on lattices available to us
we use the finite size scaling (FSS) method. Unlike previous bulk measurements
of the critical exponents, when one has to keep the correlation length
much smaller than the size of the box, in the FSS we
let the correlation length saturate at the box size. This
allows us to reach the Ginzburg region.

The results of the FSS analysis for $L_t=6$ lattices
are summarized in Table \ref{tab:fss}.
\begin{table}[hbt]
\caption{Summary of FSS results and comparison with Ising model and
MF scaling behavior.}
\label{tab:fss}
\setlength{\tabcolsep}{1pc}
\begin{tabular}{|l|lll|}
\hline
Exponents & FSS & Z(2) & MF \\
\hline
$\nu$     & 1.00(3) & 1 & 0.5 \\
$\beta_m/\nu$ & 0.12(6) & 0.125 & 1 \\
$\gamma/\nu$ & 1.66(9) & 1.75 & 2 \\
\hline
\end{tabular}
\end{table}
We see that all measured exponents show that the scaling
very close to criticality is that of the Ising model
in 2 dimensions rather than the MF one.

\subsection{The crossover exponent $x$}

\begin{figure}[t]

                \centerline{ \psfig{silent=,file=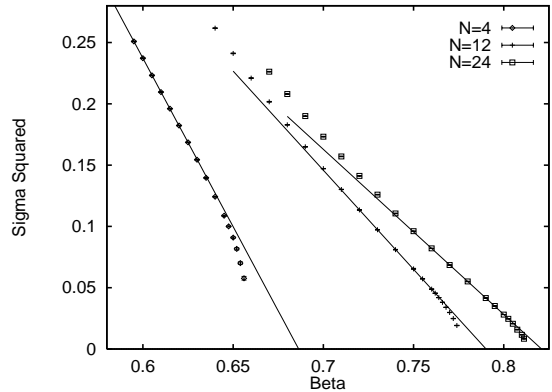,width=3in} }
\caption[]{Order parameter squared vs. $\beta$ for lattice theories
with $N=4,12,24$ ($L_t=6$).
The straight lines are the fits to the data in the MF regions.}
\label{fig:broken}
\end{figure}

A straightforward way to find the value of $x$
is to study the dependence of the
order parameter, $\Sigma$, on $\beta$. Since $\Sigma$ vanishes at the
critical point, it can be thought of as a measure of the distance from
the criticality.  We expect that for sufficiently small $\Sigma$, i.e.,
close to $\beta_c$, this dependence should be given by a power-law
scaling of the 2d Ising model:
$\Sigma\sim({\rm const}-\beta)^{1/8}$. 
 For
larger $\Sigma$, further away from the criticality, the MF scaling
holds: $\Sigma\sim({\rm const'}-\beta)^{1/2}$. 
For even larger $\Sigma$ we should see the
scaling corresponding to the fixed point of the 3d Gross-Neveu model
\cite{HaKoKo93}: $\Sigma\sim({\rm const''}-\beta)^1$.
In our simulations we can clearly resolve the MF region with the
crossover towards the 2d-Ising region (Figure \ref{fig:broken}). 

In order to find the exponent
$x$ we use the relation: $\xi=1/(2\Sigma)$ valid in the MF region.
Using the histogram reweighting method and monitoring the quality
of the linear fit as we add or subtract points we find for the
boundary of the MF region: $\Sigma_{\rm MF}=0.377(11)$,
for $N=4$, $\Sigma_{\rm MF}=0.213(4)$ for $N=12$, and 
$\Sigma_{\rm MF}=0.168(20)$ for $N=24$.
It is clear that the non-trivial 2d-Ising region is squeezed as $N$
increases. The fit $\Sigma_{\rm MF}={\rm const}\cdot N^{-x}$ gives
$x=0.51(3)$  which is in agreement with the analytical prediction
$x=0.5$ of eq.~(\ref{ginzburg}).

\section{Discussion and conclusions}

We have shown that a suppression of the critical region occurs in theories
with large number of fermion species. The corresponding crossover
exponent $x$ can be also determined (\ref{ginzburg}).

The role of the fermions is to screen the effective
self-coupling of the scalar field, $\lambda$. The strength of this
effect depends on two factors: (i) large $N$, and (ii) large window of
scales between the cutoff of the effective theory, $\Lambda$, and the
temperature, $T$.  In QCD there is
almost an order of magnitude window between the scale of the 
spontaneous symmetry breaking $\Lambda\sim 1$ GeV, and $T_c\approx
160$ MeV, which 
is presumably sufficient to drive the effective self-coupling of the
scalar field to its infrared fixed point value at the scale of $T_c$.
 How
small this value is then depends on the number of the fermions (the
condition (i)). The value $N_c=3$, though not very large, can
be considered large in some cases. One can expect, therefore, that
in QCD the width of the critical region, where actual 
O(4) exponents are observed is suppressed. Further analysis is
required to make a quantitative prediction.


\begin{thebibliography}{9}
\bibitem{PiWi84} 
		R. Pisarski and F. Wilczek, Phys. Rev. D29 (1984)
		338; K. Rajagopal and F. Wilczek,
		Nucl. Phys. B399 (1993) 395.
\bibitem{St95}
		M.A. Stephanov, Phys. Rev. D52 (1995) 3746.
\bibitem{RoSp94}
		B. Rosenstein, A.D. Speliotopoulos, and H.L. Yu, 
		Phys. Rev. D49 (1994) 6822.
\bibitem{KoKo94} 
		A. Kocic and J.Kogut, 
		Phys. Rev. Lett. 74 (1995) 3109, 
		Nucl. Phys. B455 (1995) 229.
\bibitem{KoSt98}
		J.B. Kogut, M.A. Stephanov, and C.G. Strouthos,
		hep-lat/9805023. 
\bibitem{HaKoKo93}
                S. Hands, A. Kocic, and J.B. Kogut,
                Ann. Phys. 224 (1993) 29.
\end{thebibliography}
\end{document}